\documentclass[conference]{IEEEtran}
\IEEEoverridecommandlockouts
\usepackage{cite}
\usepackage[utf8]{inputenc}
\usepackage{amsmath,amssymb,amsfonts}
\usepackage{algorithmic}
\UseRawInputEncoding
\usepackage{graphicx}
\usepackage{textcomp}
\usepackage{xcolor}
\PassOptionsToPackage{hyphens}{url}
\usepackage{hyperref}
\def\BibTeX{{\rm B\kern-.05em{\sc i\kern-.025em b}\kern-.08em
    T\kern-.1667em\lower.7ex\hbox{E}\kern-.125emX}}

\title{Insights from Railway Professionals: Rethinking Railway assumptions regarding safety and autonomy.}

\author{\IEEEauthorblockN{Josh Hunter}
\IEEEauthorblockA{\textit{University of York} \\
\textit{Centre for Assuring Autonomy }\\
York, United Kingdom \\
0000-0002-6828-974X}
\and
\IEEEauthorblockN{John McDermid}
\IEEEauthorblockA{\textit{University of York} \\
\textit {Centre for Assuring Autonomy }\\
York, United Kingdom \\
0000-0003-4745-4272}
\and
\IEEEauthorblockN{Simon Burton}
\IEEEauthorblockA{\textit{University of York} \\
\textit {Centre for Assuring Autonomy }\\
York, United Kingdom}
}
\date{May 2024}

\begin{document}

\maketitle
\begin{abstract}
This study investigates how railway professionals perceive 'safety' as a concept within rail, with the intention to help inform future technological developments within the industry. Through a series of interviews with drivers, route planners, and administrative personnel, the research explores the current state of safety practices, the potential for automation and the understanding of the railway as a system of systems. Key findings highlight a cautious attitude towards automation, a preference for assistive technologies, and a complex understanding of safety that integrates human, systematic and technological factors. The study also addresses the limitations of transferring automotive automation technologies to railways and the need for a railway-specific causation model to better evaluate and enhance safety in an evolving technological landscape. This study aims to bridge the gap between contemporary research and practical applications, contributing to the development of more effective safety metrics.
\end{abstract}
\section{Introduction}
Rail first saw deployment of autonomous systems in metros and urban rail several decades ago. However, the deployment of systems based on artificial intelligence (AI), e.g. for scene understanding, is happening rapidly in other transport modalities, notably road vehicles, leading to the question: 'What about rail?'.
The UK Rail Innovation Exhibition 2023 showed the majority of railway innovation, including the use of AI, being in the areas of sustainability, passenger experience and fuel efficiency \cite{department_for_transport_rail_2023}. So, would the introduction of technologies such as AI-based driver aids  be beneficial in train operations and, if so, how might the pace of change be accelerated?  

If railways are to make the transition towards higher levels of autonomy, especially using AI, a key issue is safety. However, to the authors' knowledge, there has not yet been a systematic exploration with train drivers and other relevant professionals regarding the use of emerging technologies and their safe integration into train operations. We seek to fill this gap through a series of interviews with current and previous railway professionals to gather their opinions, understand the technologies from a user perspective, and to explore how safety has been achieved historically to better inform the safe introduction of higher levels of autonomy.
\subsection{Context}
This study is part of the development of the SACRED methodology \cite{hunter_safety_2024} a seven-step methodology with the intention to generate safety metrics that will inform and guide the development of automated railway systems. The 
primary motivating use case for SACRED is
%SACRED methodology has a primary case study of 
a hypothetical version of the Berlin S-Bahn with 'full automation.' When discussing automation within this work, we refer to 'Fully Autonomous Operation' (FAO) as outlined by Prof Tang Tao \cite{tang_urban_2022}, \cite{tao_automation_2022}.

\begin{figure}
        \centering
        \includegraphics[width=0.49\textwidth]{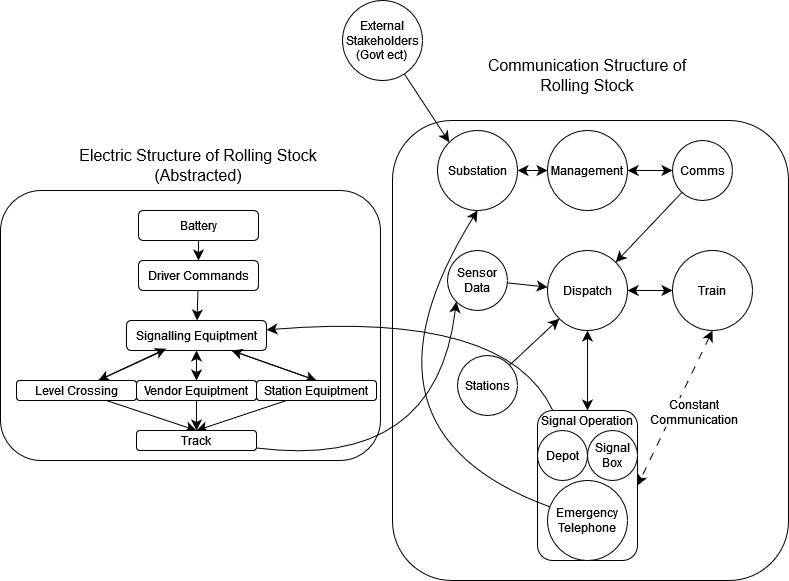}
        \caption{The Railway System of Systems model, abstracted from \cite{kaihara_system_2023} and \cite{rrpf_easy--understand_2017}. RRPF document accessed through personal communication.}
        \label{fig:SoS}
    \end{figure}

\begin{figure*}
        \centering
        \includegraphics[width=0.9\textwidth]{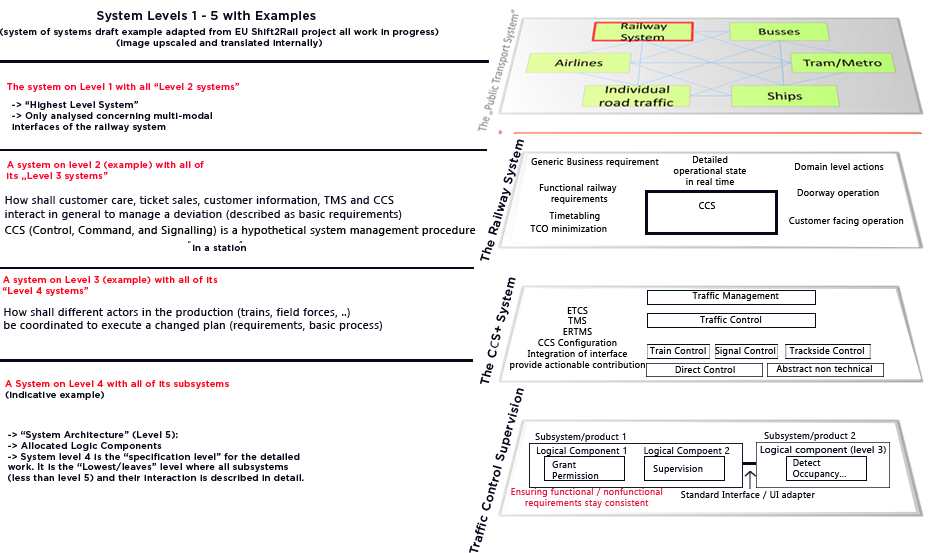}
        \caption{EU proposal for a Railway System of Systems model as proposed within \cite{rail_system_2022} Image has been modified and re-drawn for clarity.}
        \label{fig:EUSoS}
    \end{figure*}

Through this study, we contribute to the overall concept of railway as an 'ecosystem' which feeds into the idea that railway is a 'system of systems.' Decisions are made at a cab level based on information given by signalling officers, who are fed information from dispatch who gather information from a variety of sources. This relationship has been explored by Tetsuo Uzuka in the paper 'System of Systems in Railway.' \cite{kaihara_system_2023} Tetsuo argues that rail exists as two systems, each system having it's own system of systems, the electric system and the communication system, this dynamic is represented within Figure \ref{fig:SoS}.
Tetsuo's model is fundamentally focused upon the power supply, however, similar models have started to emerge from from multiple government organisations \cite{rail_system_2022, dame_weather_2021}. The EU has proposed their own 'System of Systems model' shown in Figure \ref{fig:EUSoS}.
This model assumes that 'safety' within the railway context is a well-defined, achievable goal. However, when discussing safety, metrics are particularly difficult to generate due to the low amount of incidents resulting in death or major injury, leading to multiple approaches attempting to evaluate the safety performance of railway systems \cite{sangiorgio_new_2020, korytarova_benefits_2023, blagojevic_novel_2020}. However, each of these papers arrive at similar conclusions, declaring safety to be an analysis of previous events, and defined retrospectively, meaning these approaches are difficult to apply to emerging railway systems, particularly those surrounding machine learning (ML) technology.

Traditionally, autonomous vehicles approach scoping through the generation of an Operational Design Domain (ODD). There are many definitions of what exactly an ODD should be intended to do, the specific definition we are using is PAS 1883 \textit{''The operating conditions under which a given driving automation system or feature thereof is specifically designed to function."} \cite{ccav_operational_2023} However, even within this definition, the description \textit{''Specifically designed to function"} causes issue when we consider rolling stock to be an amalgamation of pre-existing design. Other criticism of ODD focused design, such as 'the state explosion problem' \cite{valmari_state_1996} pose issue within the railway space. 

It is for these reasons that we instead propose an Operational Domain Model (ODM) approach. ODM's, as defined by Matt Osbourne and Richard Hawkins exists for systems wherein the entire operating scenario (OS) can be generalised but also the system cannot exist outside of this area. With an ODM being a list of potential hazards within a given area. So while an ODD would ask 'can a system operate within the rain, can a system operate within the dark, can a system operate while it is dark and raining' an ODM would list 'operation within rain' and 'operation within dark' as requirements through modelling \cite{osborne_defining_2023}. We use this approach due to its system agnostic nature.

\subsection{Purpose}
This study employs a qualitative approach, conducting a series of interviews with current and former railway professionals, including drivers, route planners, and administrative personnel. The interviews aim to gather comprehensive insights into their experiences and perspectives on safety, technology integration, and the broader railway ecosystem.
The purpose of this study is to understand how railway professionals perceive safety and to see if this differs from current research into the safety of driverless trains. We wish to get a greater understanding of the railway ecosystem and how to classify risk within a scenario, allowing us to create safety metrics within the space of emerging technologies. To this end, we have identified three key assumptions derived from contemporary research.

\begin{itemize}
    \item Assumption \#1  Autonomy is transferable between car and railroad
    
    Interest surrounding autonomous railway vehicles often follows innovation within the automobile industry \cite{trentesaux_autonomous_2018}. This assumes the technology is comparable and held to the same safety standards, leading to a scenario where research is done on topics that align with the technology used in automobiles, rather than railway-specific technology being created. The development of Driving Assistance technology within cars inspired the development of rail detection, autonomous stop operation and stationary object detection \cite{etxeberria-garcia_application_2020,faculty_of_engineering_computer_engineering_department_firat_university_elazig_turkey_new_2017,ye_railway_2021}.
    
    \item Assumption \#2  Railway is ready for autonomy
    
    Automation is often presented as an 'inevitability' within rail. Intuitively it would seem to be 'at least easier than the automation of a car' due to the limited operating area, limited control and predictability of the traffic. This intuitive sentiment has been argued against in previous research into the 'meaning' of autonomy and the encapsulation of the railway domain has shown the complexities regarding railway automation \cite{tonk_operational_2022, tonk_towards_2021} however this has not stopped casual/common thought regarding railway automation and is still an assumption that needs to be challenged.
    
    \item Assumption \#3 The ecosystem can properly quantify safety.
    
    This assumption posits that ’safety’ within the railway context is a well-defined, achievable goal. However, when discussing safety, metrics are particularly difficult to generate due to the low amount of incidents resulting in death or major injury, leading to multiple approaches attempting to evaluate the safety performance of railway systems \cite{blagojevic_novel_2020, korytarova_benefits_2023, sangiorgio_new_2020}. However, each of these papers arrive at similar conclusions, declaring safety to be an analysis of previous events and declared postmortem, meaning these approaches are difficult to apply to emerging railway systems, particularly those surrounding ML technology.
   
\end{itemize}
While much of the current research focuses on automating the cabin itself rather than the entire system, this research aims to investigate how our assumptions will translate to automation of the wider ecosystem.
By investigating these assumptions through the lens of professional insights, this study aims to bridge the gap between contemporary research and practical applications in railway safety. 
\section{Interview Participants}
Participants were selected through collaboration with the UK railway workers union ASLEF,  and were given questions based partially on experience and partially from a 'global question pool.' Global questions particularly revolved around objects that would be present within a railway ODM with an example being 'given your exposure to railway environments (or stories from colleagues) what is the strangest item you are aware of being found on a track?" The question is intentionally left vague in order to allow for individual interpretation, the purpose of this is to mimic previous railway safety research where the designation of problem items were left to the discretion of domain experts \cite{collart-dutilleul_towards_2019}.

The interviewees were divided into three distinct groups: Administration, Driver and Developer. Those in the Administration category included railway personnel involved in human resources, training, or safety administration. The Driver category comprises of individuals whose primary experience is operating the trains, including drivers of passenger or freight trains. The Developers category consisted of individuals who have made significant decisions affecting the track layout, including designing routes or improvements to existing routes. The overview of participants is shown in Table I:

\begin{table*}[]
\label{Table 1}
\begin{tabular}{llll}
\multicolumn{4}{l}{Table I: Participant Summary}                                                                                                                                                                                                                                                                                                                                  \\
\multicolumn{1}{|l|}{Participant ID} & \multicolumn{1}{l|}{Participant Experience}                                                                                                   & \multicolumn{1}{l|}{Participant Group} & \multicolumn{1}{l|}{Sample Question}                                                                                                              \\
\multicolumn{1}{|l|}{1}              & \multicolumn{1}{l|}{\begin{tabular}[c]{@{}l@{}}Driver (10 years) \\ Driver Training Personnel (8 years)\end{tabular}}                        & \multicolumn{1}{l|}{Administration}    & \multicolumn{1}{l|}{\begin{tabular}[c]{@{}l@{}}''What is the most common failstate\\ you see for trainee drivers?"\end{tabular}}                   \\
\multicolumn{1}{|l|}{2}              & \multicolumn{1}{l|}{\begin{tabular}[c]{@{}l@{}}Driver (30 years), Safety Engineer (10 Years, ASLEF)\\ Route Designer (Metro)\end{tabular}}  & \multicolumn{1}{l|}{Developer}         & \multicolumn{1}{l|}{\begin{tabular}[c]{@{}l@{}}''What are the standards for railway\\ route design?"\end{tabular}}                                 \\
\multicolumn{1}{|l|}{3}              & \multicolumn{1}{l|}{Driver (50 years) Retired}                                                                                                & \multicolumn{1}{l|}{Driver}            & \multicolumn{1}{l|}{\begin{tabular}[c]{@{}l@{}}''What was the most out of the\\ ordinary object you remember seeing\\ on the track?"\end{tabular}} \\
\multicolumn{1}{|l|}{4}              & \multicolumn{1}{l|}{\begin{tabular}[c]{@{}l@{}}Driver (~20 years), Safety Engineer (15 years, Corp)\\ Route Designer (Regional)\end{tabular}} & \multicolumn{1}{l|}{Developer}         & \multicolumn{1}{l|}{\begin{tabular}[c]{@{}l@{}}''When designing a new\\ route, what data points inform\\ your decisions?\end{tabular}}             \\
\multicolumn{1}{|l|}{5}              & \multicolumn{1}{l|}{Driver (2 years)}                                                                                                         & \multicolumn{1}{l|}{Driver}            & \multicolumn{1}{l|}{\begin{tabular}[c]{@{}l@{}}''What was the on boarding process\\ like for becoming a driver?"\end{tabular}}                     \\
\multicolumn{1}{|l|}{6}              & \multicolumn{1}{l|}{Freight Driver (11 years)}                                                                                                & \multicolumn{1}{l|}{Driver}            & \multicolumn{1}{l|}{\begin{tabular}[c]{@{}l@{}}''How does the experience of driving\\ freight differ from passenger?"\end{tabular}}                  \\
\multicolumn{1}{|l|}{7}              & \multicolumn{1}{l|}{Union Safety Personnel (6 years)}                                                                                         & \multicolumn{1}{l|}{Administration}    & \multicolumn{1}{l|}{\begin{tabular}[c]{@{}l@{}}''What is the largest issue you have\\ had to resolve?"\end{tabular}}                               \\
\multicolumn{1}{|l|}{8}              & \multicolumn{1}{l|}{Union Safety Personnel (7 years)}                                                                                         & \multicolumn{1}{l|}{Administration}    & \multicolumn{1}{l|}{\begin{tabular}[c]{@{}l@{}}''What is the largest issue you have\\ had to resolve?"\end{tabular}}                              
\end{tabular}
\end{table*}

\subsection{Interview Structure}
Each interview took place remotely over Zoom, lasting between 45-75 minutes, the interviews followed a set structure, each beginning with generic questions regarding the key assumptions, allowing for tangential discussion, before discussing causation,  the sample questions (see Table I) and finally ending with the question ''In a phrase, how would you define railway safety?" This structure was intended to gather comprehensive information for the purpose of this study while also allowing participants to express their insights and experiences freely, ensuring a thorough understanding of their perspectives on railway safety and automation.

\section{Findings}
\subsection{Exploring the Assumptions}
To address assumptions \#1 and \#2, all participants were asked questions regarding the future of their industry, where they see technology going and their thoughts regarding the future. Universally there was a sentiment regarding the fundamental difference between operating a train, operating a car and the attitudes towards both; trains are not like cars, with participants 7-8 (who were interviewed simultaneously) stating \textit{''While driving a car isn't easy, there's less expected of you, you are expected to react to signals as they come, whereas with the railway, you need to know the signal locations, the pre-signal types, the location of level crossings and the station entry speed kilometers before you have to react to them."} The participants went on discussing how driving a new route requires multiple weeks, occasionally months of cab riding and passing an exam. 

However, it would be incorrect to suggest that there is a complete rejection of assumption \#1. While there was little desire for removal of drivers and a rejection of the industry being ready for that, there was a positive sentiment regarding driver-assistance technology, particularly tech assisting  parsing of signs/signals. Participants discussed previous work regarding BAE Head-up displays present within aircraft. They recalled a previous study wherein drivers were enthusiastic regarding the technology coming to rail, \cite{golding_head_2008} although the study had been mostly dormant, it has recently resurfaced within the U.S Department for Transportation \cite{rzdemir_development_2022}.

\subsection{Railway as an Ecosystem}
Interviews with drivers discussed the extent to which drivers cede control to  the signal officers. Interestingly, this was the key factor which differed most from driver to driver. Every train has at least one signalling officer, a worker who typically operates within a Remote Operation Center and dictates the route a train will take and how the driver should react to hazards/anomalous objects outside of their immediate vicinity. A majority of trains have a guard, a worker who deals with immediate on-board safety hazards, things such as fires or passenger problems. The guard also has jurisdiction over the stopping of the train if they deem it necessary. The railway driver handbook suggests that drivers have the right to override the commands of the guard or signalling officer with 'good enough reason.' \cite{rssb_preparation_2023} The information from the handbook and the communication with the drivers suggests that the 'railway ecosystem' is defined as the relationship between the driver, the signal officer and the guards on individual routes and all potential routes the train could be diverted towards.

However, discussion with the other participant groups revealed an ecosystem with a much wider scope. While all participants were either drivers or driver adjacent, some insight was given on the ecosystem outside of the cab. Participant \#7 stated:

\textit{''The largest common problem we face (as Union Safety Personnel) is anything that cannot be dealt with locally, if a driver has to go 4 hours in-between stops and needs the bathroom half way, are they going to be able to focus on the track? If a train has to drive over a 120 year old embankment each day, how often is that getting checked?"}

The specific incident the participant was making reference towards was the Stonehaven derailment, \cite{raib_derailment_2020} which occurred after rolling stock travelled over debris caused by old earthworks within the area. The issues with the earthworks had been heavily documented over time and a resolution was in place in the form of a debris drainage system. The drain, however, had not been designed to code; further comment is available within the RAIB report on the matter \cite{network_rail_digital_2019}. Overall, this speaks to the relation of the ecosystem. Due to the shape of the landscape and quality of signalling information at the location, early detection of the rubble seems impractical and instead of being an issue for any cab-adjacent personnel to resolve, this issue in particular serves to highlight the importance of the wider ecosystem, the environmental agencies, meteorologists and other such personnel whom all are part of the 'railway ecosystem.' Furthermore, participants \#7 and \#8 discussed what they called 'generic safety' in detail, discussing how the comfort of cab personnel through the form of regular bathroom breaks/cab temperatures are constantly at risk and a safety concern.

In conclusion, the railway ecosystem includes all systems that have influence over the operation of the train. The primary system being the cab itself, together with signal engineers, drivers and guards. Additional systems include the railway environment,  meteorologists, track designers and environmental agencies, and the systems behind route planning, rolling stock development and driver training. The railway ecosystem is complex, and discussing it in full is outside of the scope of this work, however, a key finding from our discussions is that introducing autonomy requires a broad understanding of the ecosystem, not just a focus on in-cab technologies.

\subsection{Definition of Safety}
Every interview ended with the question, ''How would you describe safety?" This question revealed an interesting pattern. Drivers perceived safety as almost predetermined, summarizing it as\textit{''the ability to get from A to B without hitting anyone."} Participant \#3 states 

\textit{''The line itself is safe, everything is there for you to do your job safely, you just need to make sure you do it right and don't mess anything up."} 

This response reflects a significant trust in the railway ecosystem as a whole. When discussing safety with developers, their responses were similarly systematic, with an attitude of,  \textit{''Safety is when all of the rules are followed."} 'Developers were particularly interested in previous cases reported by the RAIB, where the outcome of an investigation often led to a series of micro-adjustments to the rules following an incident. This suggests that safety is not a fixed goal but rather an ideal to strive for continuously.

Administrative workers, who spent the most time dealing with systematic safety, did not present a consensus. Participant \#7 discussed individual risk on any given day and how external factors (weather, traffic, infrastructure age) can affect risk. Participant \#7 also mentioned:

\textit{''There's a responsibility to keep the rail open, because if someone doesn't take the train, they're going to take the car and that means overall there's more risk that day. The risk of closing the train isn't discussed but I think it needs to be.} 

Participant \#1 discussed safety on a more systematic level, stating that ''safety" overall is a range determined by infrastructure, and the input of the driver decides where exactly on that range "safety" falls. This view is similar to the idea presented by Participant \#3 but they were more critical of infrastructure and logistics, stating: 

\textit{''We know how to do safety, the issue is when people try to cut corners to save on costs."} 

Participant \#8 did not feel confident to answer 'safety as a whole' and instead opted to say \textit{It's complicated and I'll leave it at that.}
The attitude of both drivers and developers is much like that within published research and industry reports \cite{spark_measuring_2019, dunelm_dunelm_2016, michael_woods_application_2011}, suggesting safety to be measured retrospectively  meaning as a summary of previous events, infrastructure is declared 'safe' after extended operation results in low casualties, drivers feel confident in the overall safety of their work due to it's success. This is difficult to abstract, a new rail line cannot be declared 'safe' using this approach, within the interviews with participants 2 / 4, individuals who have worked on opening new rail lines in the past were asked questions regarding assuring the safety of new lines, the answer of which can be summarised as 

\textit{'Following guidelines created for previous routes, guidelines which are updated in case of a structural incident.'} 

This response aligns with the retrospective strategy. However, administrative participants seemed to reject Assumption \#3, suggesting that safety is not an achievable goal but rather something to strive towards.

\subsection{The Ego Vehicle}
During final question 'in a sentence, how would you define railway safety' as well as the general attitude throughout the interview participants showed that within rail there exists this idea that the entire route exists as one ecosystem. Great trust is given to other individuals within the 'safety network', constant communication takes place between drivers, signal officers and guards when an incident occurs, and great care is taken to ensure blame is not placed on any individual should an incident occur. Interviewees cited cases by the Railway Accident Investigation Branch (RAIB) and the Railway Safety and Standards Board (RSSB), both independent bodies within the UK that investigate and report on safety incidents within the railway space. The RAIB have a standard philosophy of not allocating blame, stating:

\textit{The purpose of a Rail Accident Investigation Branch (RAIB) investigation is to improve railway safety by preventing future railway accidents or by mitigating their consequences. It is not the purpose of such an investigation to establish blame or liability.} \cite{raib_report_2023} 
 
This creates an interesting situation where, although in the moment decisions are ultimately made on an individual level, the outcome is dependent on the ecosystem at large, making it difficult to build any sort of metric-based safety approach that isn't retrospective. Overall, this suggests that if we wish to make further innovation within the space of railway safety, it may be prudent to look towards an exploration of this ecosystem as a large 'system of systems'. This ideology was present within the interviews of participants \#1 and \#3, and there is a general understanding that the work done by drivers is but one part of the whole when it comes to safety. This 'safety ideology' is reflected in present research, e.g. the work of Burton and McDermid around the holistic perspective of safety wherein the safety of any subsystem is directly related to the complexity of the system as a whole \cite{burton_safety_2021}. For our purposes, it suggests that rolling stock is an ego-vehicle inside a self-contained system of systems and this contrasts with autonomy for road vehicles, which coined the term ego vehicle, that takes a highly vehicle-centric view.

\subsection{Causation and Safety Metrics}
When discussing safety metrics with participants, particularly those in administration, two classifications were brought up. Signal Passed At Danger (SPAD) and 'crashworthiness.' According to the RSSB, a SPAD occurs when a train passes a signal showing a stop aspect or exceeding a movement authority without authorisation. From this, we can extract two key metrics, SPAD rate and Signal Detection Time (SDT). What is meant by 'crashworthiness' is the survival rate during post-incident reaction, the crumple zones, impact resistance and rigidity around passengers. This is a variable reported by the Office for Rail and Road (ORR). For the purposes of this study, the crashworthiness of a train is not assumed to be a factor in the introduction of autonomy. 

Interviewees suggested that SPAD risk is solved and that driver training is rigorous to the point of the limit in human capability, a sentiment echoed by Chris Harrison et al in the paper 'At the limit? Using operational data to estimate train driver human reliability.' \cite{harrison_at_2022} Harrison finds that human drivers have a SPAD rate of 1/43,000 within passenger rail and suggests that the only way to improve this figure is with technological systems such as the European Train Control System (ETCS.) \cite{janhsen_formal_1997} ETCS is an alternative method for classifying autonomy, similar to FAO.  This does seem to align with assumption \#3, however, an issue with the SPAD model as explored within ETCS is the inherent reliance upon Event-based Accident Models \cite{kim_closer_2019}. As mentioned, participants rejected any notion of a causation model which allocates blame. While the participants do not represent railway personnel as a whole, it is important to note that within the United Kingdom, ETCS is yet to be implemented to a level which would allow for any allocation of causal responsibility between human and technology. Network Rail has expressed interest in further deployment of ETCS, however they have identified the need for further research into 'digital railways.' \cite{network_rail_digital_2019}

\section{Discussion of Assumptions}
During the interviews, the assumptions were directly referenced multiple times. However, we are now taking the opportunity to reevaluate these assumptions as a whole.
\subsection{Technological Overlap}
Assumption \#1 compares the technology requirement for cars and railway. As discussed, the core differences between cars and railway are the ecosystem dynamics and the safety requirements, however, it would be incorrect to suggest that there is no overlap. The technology used within cars has inspired work within the railway space to contribute to the partial autonomy found within rail at the moment \cite{etxeberria-garcia_application_2020,faculty_of_engineering_computer_engineering_department_firat_university_elazig_turkey_new_2017,ye_railway_2021}. Although this work has been framed with the goal of 'total automation' it would be incorrect to suggest that this assumption is false, as partial automation of the railway has been common since the 1920's \cite{liu_unmanned_2021}. This partial automation is not without it's safety issues, which we have investigated in previous work \cite{collart-dutilleul_investigating_2022}, and the transition to 'total automation' is proving to be an issue. However, when discussing this assumption in the interviews, responses were mixed. One conclusion that can be drawn is that the technology used within automotive automation are useful tools for helping to guide railway automation, however a direct translation is not possible, and technology will need to be adapted for rail, and wider technological solutions will be needed for the automation of everything outside of the cab. 

Currently existing railway systems with technology that can be classified as 'full autonomy' are either highly restricted domains in the case of the Paris-1 line \cite{chatelus_paris_2014} or with the automation of a new, small, ecosystem in the form of the the Docklands Light Rail (DLR). Interestingly, participants generally believed the London DLR to be 'full automation' with Participant \#5 citing the DLR as 'the end-goal of automation.' This, somewhat anecdotally is also echoed within casual understanding, when discussing automation of railways in conversation, the DLR is often an example given of 'working automation' as it is a self-contained ecosystem with driverless trains operating during all hours. However, investigation has revealed that the Transport for London (TfL) board does not recognize DLR as GoA-4 or any form of 'high automation' as classified within our scope \cite{tfl_foi_2024}. Somewhat interestingly the TfL also notes no interest in improving the Grades of Automation within the ecosystem of the London underground, suggesting a rejection of assumption \#2. Overall, it can be suggested that the automation of an entire 'railway ecosystem' has not yet been formally attempted outside of closed, single track ecosystems such as the Paris-1 line. 

\subsection{Automation of an Ecosystem}
\begin{figure}
    \centering
    \includegraphics[width=0.49\textwidth]{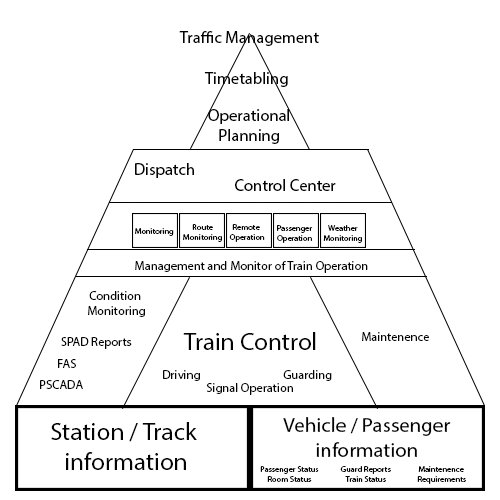}
    \caption{'Fully Autonomous Operation' traffic management as defined by Professor Tang Tao \cite{tao_automation_2022}. Translated and documented via internal communications.}
    \label{fig:FAO Eng}
\end{figure}

When discussing assumption \#2, it is important to discuss the complexity of not only the 'system of systems' but each system itself. Within this paper we have discussed the systems of 'cab', 'environment' and 'systems planning'. It is interesting to note that the typical time it takes a driver to learn their first route is between 13-18 months. This need for extensive and intimate knowledge places a significant technological demand on automating even a single part of one system. While this does not make automation impossible, it does suggest that efforts might be more effectively directed towards examining the ecosystem as a whole. This is a sentiment we are starting to see throughout the industry. As mentioned previously, the network rail environmental task force has started a call for cross industry collaboration \cite{dame_weather_2021} stating that rail as it exists right now risks fragmentation and has a disconnect with the state of the art. The EU has moved towards categorising the 'system of systems' into manageable groups, which they call the 'pillars of system safety', in order to tackle the problem of fragmentation \cite{rail_system_2022}. This somewhat mimics the methodology seen within the automation of railways in China through the FAO system. The FAO system \cite{tao_automation_2022} categorises the railway 'system of systems' into a 'traffic management pyramid' separating management, train control and condition monitoring, showing how each system interacts with one-another and discusses how each new technological implementation influences the system as a whole, as shown within Figure \ref{fig:FAO Eng}. Tao's FAO pyramid shares similar philosophies with the 'System of Systems' approach we are beginning to see within the EU and also somewhat mimics the philosophies held by Participants \#1, \#3 and \#7.

However, much current work reflects the premise of the train as an ego-vehicle. Work surrounding split-second decisions, such as path determination \cite{laurent_train_2024} and Horn Performance \cite{tagiew_mainline_2023} have made significant progress in the development of railway technologies with specific arguments being made for traditional ODD oriented ego-vehicle safety within rail \cite{weiss_approach_2024}. We do not believe this approach to be incorrect; the innovation made within this space is important for the development of the safety of railways and the interview participants seemed to agree. There was an understanding that 'car-like technology' could be applied to railway in order to improve the quality of life, as discussed, participants seemed enthusiastic regarding 'BAE Head-up displays' and similar technology, technology which considers the cab alone.

The drivers seem to present a holistic ideology of safety, similar to that found within Burton's previous work. What this means for railway development, particularly that regarding automation and relating to the SACRED methodology, is that work must be framed as a part of a whole, a driver extracts information themselves and make judgements in the moment, however they also heavily rely upon information from outside sources. Overall this is not a rejection of assumptions \#1 and \#2, but rather a suggestion that the railway ecosystem is not so simple that road vehicle technology can be transferred to rail as they exist right now.

What we can conclude from reading the surrounding research as well as our communication with the professionals is that more work must be done to generate cohesion between disciplines and to further understand the overall 'system of systems' that is rail. However, the system of the cab itself has a healthy amount of research in the area and the technology seems promising to industry professionals. For our purposes, generating safety metrics and verifying our assumptions, we agree with the researchers at Network Rail that rail must do more to overcome fragmentation within the industry.

\subsection{A Railway Specific Causation Model}

Assumption \#3 introduces wider problems about how railway 'safety' is considered. Interview responses resulted in a trend emerging of each participant intuitively following the hierarchy of causal influences, a model used within construction  that explores the idea of individual performance shaping an overall concept of safety rather than being a root cause of an incident \cite{haslam_contributing_2005}. This idea is consistent outside of construction with work by Svedung, I., \& Rasmussen, J. (2002) \cite{svedung_graphic_2002} which suggests that while accidents are caused by a 'critical event' these events are in turn caused by systematic failures. This, in turn, is consistent with the ideas present within RAIB reports \cite{raib_report_2023}. The idea that 'human error' is a low-factor and not significant enough to be a classifier of 'blame.' Each participant had differing levels of confidence in the idea of a single 'Railway ecosystem' rather than the rolling stock being an ego vehicle.

The Swiss Cheese model of causation seems to fit, a model which presents incident causation as a series of defects in organizational safety resulting in many small ''unsafe acts" ultimately causing one failure \cite{reason_contribution_1990}. However, Reason's model suggests that technological failures can be eliminated with technological advancement; resulting in human error being the sole remaining immediate cause of accidents \cite{hosseinian_major_2012}. For both the transition to autonomy and also to fit the current railway philosophy as presented by interview participants, it would seem correct to avoid a model which seems to allocate 'blame' but rather one that attempts to investigate incidents similarly to the RSSB/RAIB. From this, it is reasonable to suggest that typical causation models are not applicable to rail and rail itself requires its own model of causation - or at least a new model is needed where autonomous capabilities are introduced in a complex system of systems. 

\section{Conclusion}
This work provides valuable insights into the railway domain, contributing to the ongoing debate between viewing railway systems as ego-vehicles versus systems of systems. Our findings indicate that railway personnel predominantly align with the system of systems perspective. However, drivers, in particular, recognize the need for ego-vehicle-like technologies within the cab. This research highlights the significance of 'blame' within the railway context and supports the development of a railway-specific, or a complex system of systems autonomy, causation model for the future.

Other studies have identified a \textit{''Disconnect with the state of the art"} \cite{dame_weather_2021} wherein existing technologies are not readily applicable within their intended contexts. We believe this disconnect stems from differing philosophies and an over-reliance on assumptions present, or implied, in current literature. As technology advances and railway automation becomes more prevalent, it is crucial to consider how these technologies can integrate into the broader ecosystem.

One issue is the lack of consensus within the global railway community on defining a railway ecosystem. In this paper, we have discussed various approaches, including the Japanese 'System of Systems,' the Chinese 'Fully Autonomous Operation,' the European 'Pillars of System Safety,' and the ETCS/ERTMS frameworks. The railway personnel we interviewed had their interpretations of a 'railway ecosystem.' Still, there was no standard protocol followed. Drivers were familiar with technologies like ATO/ATP and understood their reliance on signal operators and other external individuals, reflecting more of an 'ecosystem philosophy' rather than strict adherence to any particular methodology.

These insights pave the way for further research into developing more effective and context-specific safety models for the railway industry. The interview findings indicate a lack of a suitable and versatile accident causation model within the diverse railway ecosystem, similar to the model used in construction as presented by Haslam \cite{haslam_contributing_2005}. However, due to the unique characteristics of the railway system, Haslam's model cannot be directly applied. Instead, this highlights the need and opportunity to develop a railway-specific causation model. This sentiment is echoed in the work by Hong et al,\cite{hong_railway_2023} underscoring the importance of tailored approaches to safety in the evolving landscape of railway automation. 

This work highlights a philosophy of safety held by professionals within the industry and underscores the need for future efforts to focus on creating a railway-specific causation model. It introduces several key issues with railway safety and suggests that the quantity of research in this area is lacking. Paradoxically, while a railway-specific causation model is needed, it would be premature at this stage due to the limited amount of work currently investigating the railway environment/ecosystem, as railway safety lags behind its counterparts in other transportation domains. This relates back to the inherent difficulty in specifying railway safety requirements, as participant \#8 said when asked the question, ''How would you define railway safety?"

\textit{''It's complicated and I'll leave it at that."} 

However, we do not intend to leave it at that. Relating these findings back to our surrounding work and the SACRED methodology, we believe it is possible to 'de-mystify' safety. SACRED's strength lies in its input-agnostic nature. Instead of classifying relationships between systems, SACRED focuses on generating safety metrics specific to a given domain through the usage of the Operational Domain Model (ODM), as opposed to the current industry standard of using an Operational Design Domain (ODD).

The insights from this study emphasize the necessity of developing safety metrics tailored to the unique aspects of the railway ecosystem, rather than relying solely on pre-existing assumptions and technologies. This sentiment is echoed within the different 'system of systems' approaches explored by various government organizations. While the assumptions and current technology have merit, generating widely applicable railway metrics requires a new perspective. It is crucial to acknowledge that any metrics generated will inherently depend on external stakeholders and their interactions, rather than being generated solely from the rolling stock as an ego-vehicle.

Regardless of philosophy, it is clear that there is a disconnect between 'state of the art' technology, practical applications, and professional perspectives. We are using the insights gained from this study to help develop our SACRED methodology and to argue for both more work within the area of Domain Modelling and the adoption of the 'System of Systems' model present in other countries.

\section*{Acknowledgment}
This work is undertaken with support from Siemens mobility, and in conjunction with the the Fraunhofer Institute for Cognitive Systems (IKS) in Munich. This work was undertaken with the cooperation of the Associated Society of Locomotive Engineers and Firemen (ASLEF).

For the purpose of open access, the author has applied a Creative Commons Attribution (CC BY) license to any Author Accepted Manuscript version arising.

Copyright 2025 IEEE. Personal use of this material is permitted.
\bibliographystyle{splncs04}
\bibliography{references}
\end{document}